\documentclass[lettersize,journal]{IEEEtran}
\usepackage{amsmath,amsfonts}
\usepackage{algorithmic}
\usepackage{algorithm}
\usepackage{array}
\usepackage[caption=false,font=normalsize,labelfont=sf,textfont=sf]{subfig}
\usepackage{textcomp}
\usepackage{stfloats}
\usepackage{multirow}
\usepackage{url}
\usepackage{verbatim}
\usepackage{graphicx}
\usepackage{cite}
\usepackage{tabularx}
\usepackage{makecell}
\usepackage{xcolor}
\usepackage{hyperref}
\begin{document}

\title{A Computational Procedure for Assessing I$_c$($\varepsilon$) in Nb$_3$Sn/Bi-2212 Hybrid Magnets}

\author{A. D’Agliano, A. V. Zlobin, I. Novitski, G. Vallone, P. Ferracin, E. Barzi,~\IEEEmembership{Senior Member,~IEEE}, S. Donati and V. Giusti
\thanks{This work was supported in part by the U.S. Department of Energy, in part by the Office of Science, in part by the Office of High Energy Physics, through the Magnet Development Program, under Contract DE-AC02-05CH11231, which is performed at Lawrence Berkeley National Laboratory.
This work was also supported by the EU Horizon 2020 Research and Innovation Program under the Marie Sklodowska-Curie Grant Agreement No. 734303,822185, 858199, and 101003460, and the Horizon Europe Research and Innovation Program under the Marie Sklodowska-Curie Grant Agreement No. 101081478.

A. D'Agliano is with the Lawrence Berkeley National Laboratory, Berkeley, CA, 94720, USA, and also with the University of Pisa, Pisa, 56126, Italy (Corresponding author, e-mail: adagliano@lbl.gov).

I. Novitski, and A. V. Zlobin are with the Fermi National Accelerator Laboratory, Batavia, IL, 60510, USA.

G. Vallone and P. Ferracin are with the Lawrence Berkeley National Laboratory, Berkeley, CA, 94720, USA.

E. Barzi, is with Ohio State University, Culumbus, OH, 43210, USA.

S. Donati and V. Giusti are with the University of Pisa, Pisa, 56126, Italy.}

\thanks{Manuscript received July 28, 2025; revised September 12, 2025; accepted October 15, 2025}}

\markboth{IEEE Transactions on applied superconductivity,~Vol.~X, No.~X, XX}%
{Shell \MakeLowercase{\textit{et al.}}: A Sample Article Using IEEEtran.cls for IEEE Journals}

\IEEEpubid{0000--0000/00\$00.00~\copyright~2025 IEEE}

\maketitle

\begin{abstract}
The critical current of superconductors is commonly measured by testing unloaded wires under an external magnetic field. While stressed by intense Lorentz forces, the existing HTS/LTS superconductors are prone to a reduction in critical current before reaching their structural mechanical limit.
In this work, the magnetic and mechanical analysis of the FNAL 4-layer Bi-2212/Nb$_3$Sn hybrid dipole magnet is reported, aimed at predicting the critical current degradation for both the superconductors during powering at 16 T. All the Rutherford cables in the coils of the hybrid magnet were modeled at the strand level in Ansys APDL with the heterogeneous cable model.
Utilizing this detailed geometry, it was possible to evaluate the effects of strain on the critical current degradation for both the Nb$_3$Sn and Bi-2212 superconductors under the intense Lorentz forces. The analysis presented in this paper integrates strain-dependent critical current laws, with parameters derived from experimental data, to simulate the hybrid magnet's performance for all possible current-powering configurations. The proposed methodology enables a detailed assessment of conductor integrity and I$_C$($\varepsilon$) reduction in existing hybrid magnet designs, providing a versatile and rigorous framework for optimizing future high-field hybrid magnets.


\end{abstract}


\section{Introduction}

\begin{figure}
\centering
\includegraphics[width=2.8in]{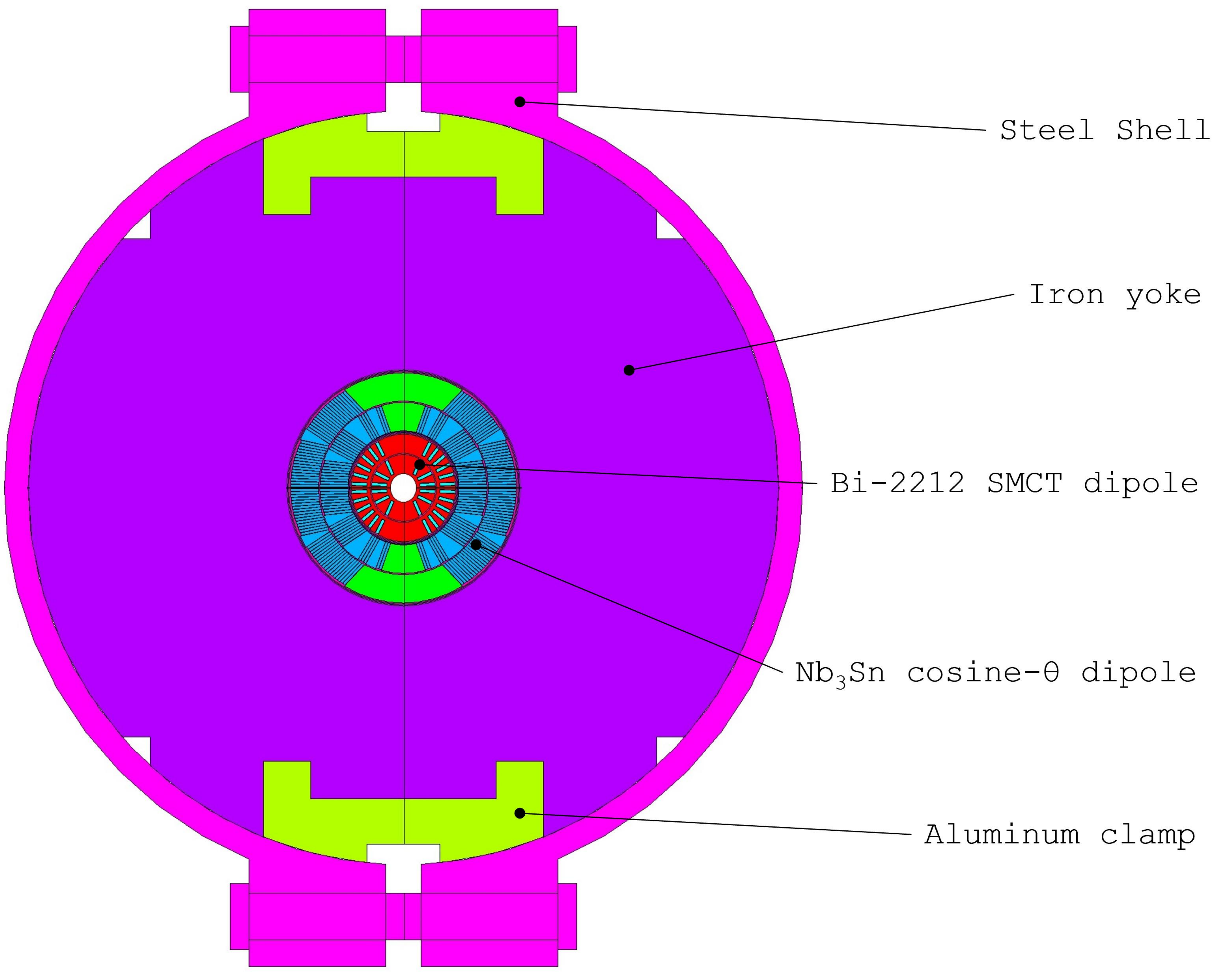}
\caption{Schematic representation of the 4-layer hybrid magnet in dipole configuration, with the insert coil made of Bi-2212 Rutherford cable (Bi-SMCT1) and the outer cosine-theta coil in Nb$_3$Sn.}
\label{fig:hybrepr}
\end{figure}

\IEEEPARstart{T}{he} main goal of Area II of the U.S. Magnet Development Program (MDP) is to demonstrate the performance of HTS magnets with the intent of developing and testing hybrid magnets beyond the limit of 16 T, advancing the technology of high and low temperature superconductor (HTS/LTS) \cite{USMDP}. At high field, the stress and strain state inside the superconductor strands can reach critical levels due to intense Lorentz forces, particularly in the innermost layers of the hybrid magnet. Such mechanical loads can degrade the performance of both HTS and LTS superconductors. \IEEEpubidadjcol 
To control the distribution of stress and strain within the conductors, innovative tools and technology have been developed: from the design perspective, the structural geometry called stress-managed cosine-theta (SMCT) has been introduced to control the stress intensity in the conductor \cite{LTS1, LTS2, LTS3, LTS4, LTS5}. From a computational perspective, detailed cable models at the strand level, known as heterogeneous models, have been previously developed to assess conductor integrity and evaluate the reduction in critical current caused by strain \cite{GVmethod, ADmodel, CCDGiorgio, FNALmodel, Irreversible}.
In this paper, we present the magnetic and mechanical analysis of the FNAL HTS/LTS dipole by applying a computational procedure that implements the heterogeneous Rutherford cable model for the first time to the entire hybrid magnet, cross-section in Figure \ref{fig:hybrepr}. The critical current degradation law due to strain for Nb$_3$Sn was implemented as described in \cite{CCDBordini}. For the Bi-2212 superconductor, the degradation law due to strain was obtained from the transverse pressure I$_C$ degradation measurements performed by the University of Twente \cite{CCDBi2212}. The strain-state and magnetic field results from the FEM analysis in each superconductor region were then used as input to compute the I$_C$(B, T, $\varepsilon$) reduction across all the possible current powering configurations of the hybrid dipole.
This approach allowed the detailed analysis of an existing hybrid magnet design.
The parametric nature of this procedure enables its application to different or new dipole geometries, providing a valuable tool for scientists and engineers involved in the design and optimization of future high-field hybrid magnets.

\section{Methodology}

\subsection{Heterogeneous Rutherford cable model}
\begin{figure}
\centering
\includegraphics[width=3.3in]{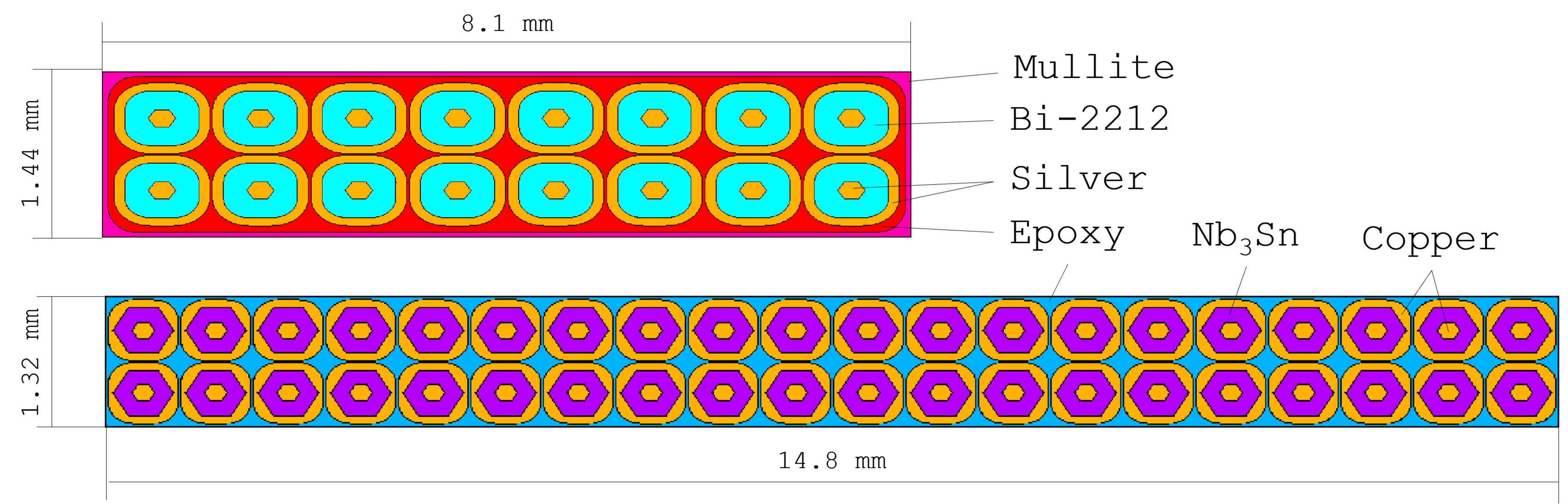}
\caption{Cross-section of the heterogeneous Rutherford cable model for the Bi-2212 (top) and Nb$_3$Sn (bottom) superconductors (cables are not in scale).}
\label{fig:RCs}
\end{figure}

Rutherford cable models, detailed at the strand level, were implemented in ANSYS APDL to obtain a precise stress and strain state distribution in the magnet coils and to compute the accurate critical current reduction due to strain during powering. Table \ref{tab:bicabpar} reports the strand and Rutherford cable parameters for the Bi-2212 and Nb$_3$Sn superconductors. Figure \ref{fig:RCs} shows the materials and the geometrical dimensions used as input in ANSYS APDL to implement the heterogeneous Rutherford cable model for the first time on a hybrid magnet design. Material properties for the Nb$_3$Sn cable could be found in \cite{Areview}, while those for the Bi-2212 cable are reported in Table \ref{tab:matprop}.

\begin{table}[!t]
\centering
\caption{Rutherford cables and strands parameters\label{tab:bicabpar}}
\begin{tabular}{p{5.5cm} p{1.2cm} p{.9cm}}  
\hline
 \textbf{Parameter} & \textbf{values} & \textbf{unit} \\
 \hline
 \textbf{Bi-2212 LBNL-1110 Cable, PMM180207-2 Strand} & & \\ \hline
 Number of strands & 17 &  \\
 Bare cable width & 7.8 & mm \\
 Bare cable thickness & 1.44 & mm \\
 Strand diameter before/after reaction & 0.80/0.778 & mm \\
 Strand J$_C$ (4.2 K, 5 T) & 460 - 640 & A/mm$^2$ \\
 \hline
 \textbf{Nb$_3$Sn cable, RRP-108/127 Strand} &  &  \\ \hline
 Number of strands & 40 &  \\ 
 Bare cable width & 14.847 & mm \\ 
 Bare cable mid-thickness & 1.307 & mm \\ 
 Strand Diameter & 0.7$\pm$0.003 & mm \\
 Strand J$_C$ (4.2K, 12T) & 2643 & A/mm$^2$ \\
  \hline
\end{tabular}
\end{table}

\begin{table}[bt]
\centering
\caption{Material Properties at 300/4.2 K\label{tab:matprop}}
\begin{tabular}{p{2.3cm}  c c c}  
\hline
\multirow{4}{*}{\thead{\textbf{} \\ \textbf{Material}}} & \multicolumn{3}{c}{\thead{\textbf{Properties value}}} \\
\cline{2-4}
                                                            &
\thead{\textbf{Young} \\ \textbf{Modulus} \\ \textbf{[GPa]}} &
\thead{\textbf{Poisson} \\ \textbf{Ratio}} &
\thead{\textbf{Yield} \\ \textbf{Strength} \\ \textbf{[MPa]}} \\
\hline
Silver & 83/91 &  0.364  & 37/52 \\ 
Bi-2212 & 45/50 & 0.35 & 37/52 \\ 
Mullite & 13/20 & 0.31 & 500/600 \\ 
Epoxy & 5/6 & 0.39  & 87/96 \\ 
\hline
\end{tabular}
\end{table}

\subsection{Critical Current Scaling Law - Nb$_3$Sn}\label{sec:CCD}
The method used to define the Nb$_3$Sn critical current as a function of the applied magnetic field and the strain-state within the superconductor strands at 4.2 K was obtained from \cite{CCDBordini}. To evaluate the I$_C$($B,T,\varepsilon$), the critical surface of the 40-strand Nb$_3$Sn Rutherford cable was parametrized. Then the 'strain' function s($\varepsilon$) was added to the I$_C$ equation using the exponential scaling law and assuming: C$_1 = 0.89$ and $\varepsilon_{l0} = -0.2$. The strain function was computed as:

\begin{equation}
    s(\varepsilon) = \frac{B_{c2}(0,\varepsilon)}{B_{c2}(0,0)} = \frac{e^{-C_1 \frac{J_2 + 3}{J_2 + 1} J_2} + e^{-C_1 \frac{I_1^2 + 3}{I_1^2 + 1} I_1^2}}{2}
\end{equation}



where $I_1$ and $J_2$ are respectively the first and second invariants of the area-averaged strain tensor $\varepsilon$ computed in each Rutherford cable's strand. The strain function can be defined as the ratio of the strain-dependent upper critical field B$_{c2}$ at zero temperature, and B$_{c2}$ at zero temperature and strain. The I$_C$(B, T, $\varepsilon$) depends but is not proportional to the strain function s($\varepsilon$).

\subsection{Critical Current Degradation Law - Bi-2212}

\begin{figure}
\centering
\includegraphics[width=2.4in]{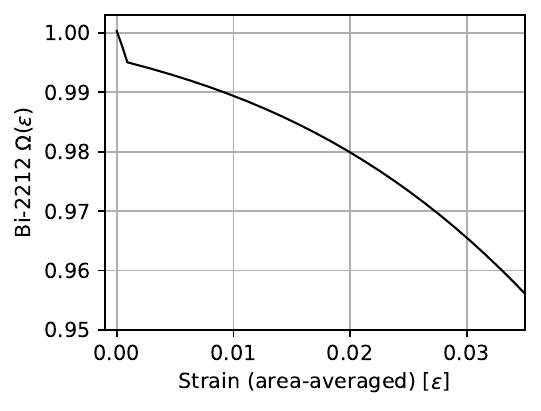}
\caption{Normalized I$_C$ degradation $\Omega$($\varepsilon$) (at 4.2 K, 11 T) as a function of the area-averaged strain state in the Bi-2212 Rutherford cable strands due to transverse pressure.}
\label{fig:CCDBi}
\end{figure}

The critical current degradation of the Bi-2212 conductor due to transverse pressure was obtained from recent measurements at Twente University \cite{CCDBi2212}. A mechanical FEM simulation reproducing Twente's experimental setup of a single Bi-2212 Rutherford cable was performed to correlate the azimuthal pressure with the strain state inside the cable's strands. Combining those two relations, the critical current degradation law due to strain in the HTS Rutherford cable was obtained, as shown in Figure \ref{fig:CCDBi}. The initial steep trend of the relative I$_C$ is associated with the cable's elastic regime. When the Bi-2212 composite cable reaches the plastic regime, the degradation gradient drastically reduces.

The relative critical current can be interpreted as a scaling coefficient that reduces the nominal I$_C$ due to the strain state intensity. For that reason, in this work, it was defined as the ratio of the degraded critical current and the critical current at zero strain: $\Omega$($\varepsilon$) = I$_C$(B, T, $\varepsilon$) / I$_C$(B, T, 0). In this formulation, $\Omega$($\varepsilon$) is independent of the magnetic field in the conductor.

\section{Magnet Design}
\subsection{Magnetic assumptions}

\begin{table}
\centering
\caption{4-Layer Hybrid Dipole Configuration \label{tab:ngeomelepar}}
\begin{tabular}{p{5.5cm} c}
\hline
\textbf{Parameter} & \textbf{values} \\
\hline
Bi-2212 Insert ID - OD, mm & 16 - 58 \\ 
Nb$_3$Sn Outsert ID - OD, mm & 60 - 122 \\
Number of Bi-2212 turns, L1 - L2 & 3 - 6 \\
Number of Nb$_3$Sn turns, L1 - L2 & 22 - 34 \\
Iron yoke OD, mm & 400  \\
Max. transverse size w/ shell, mm & 545 \\ \hline
\end{tabular}
\end{table}

\begin{table}
\centering
\caption{Hybrid Dipole Magnetic Results at 4.2 K\label{tab:magres}}
\begin{tabular}{p{3.5cm} c c}  
\Xhline{3\arrayrulewidth}
\multirow{2}{*}{\thead{\textbf{}}} &       \multicolumn{2}{c}{\thead{\textbf{Powering}}}     \\
\cline{2-3}
                                            & \thead{\textbf{Series}} & \thead{\textbf{Optimal}} \\
\hline
I$_{HTS}$ [kA] &  9.6  &  9 \\
I$_{LTS}$ [kA] &  9.6  &  13.5 \\
Bore field [T] &  12.9  &  15.9 \\
B conductor, L1 [T] &  13.4  &  16.3 \\
B conductor, L2 [T] &  12.4  &  15.4 \\
B conductor, L3 [T] &  11.4  &  14.7 \\
B conductor, L4 [T] &  9.5  &  12.3 \\
\Xhline{3\arrayrulewidth}
\end{tabular}
\end{table}

\begin{figure}
\centering
\includegraphics[width=3.5in]{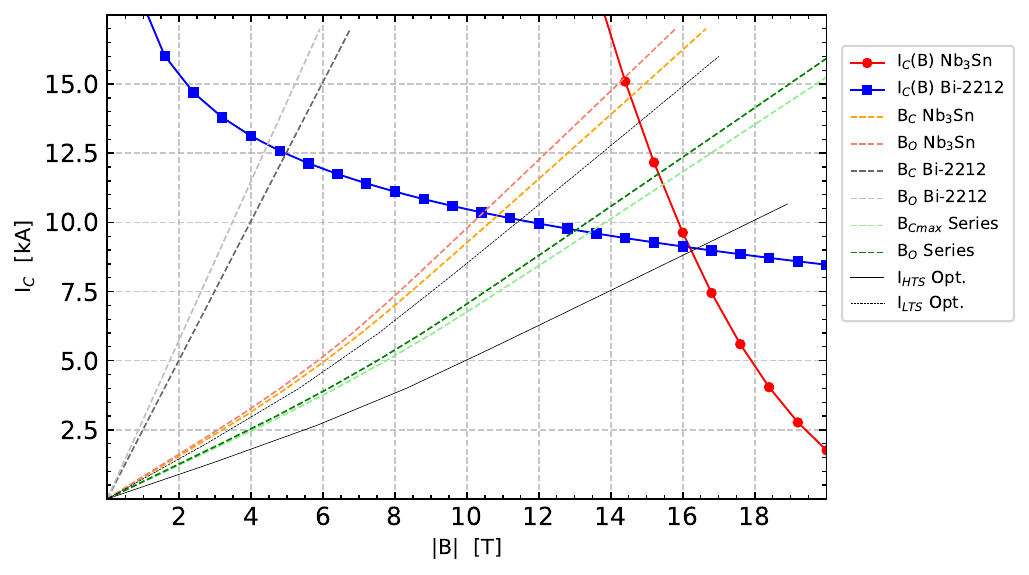}
\caption{I$_C$(B) curves of the Bi-2212 (blue) and Nb$_3$Sn (red) superconductors at 4.2 K with the load-lines of: powering HTS insert only (grey), powering LTS outsert only (orange), hybrid dipole in series powering configuration (green), and optimal load-lines for both HTS and LTS conductors (thin black lines).}
\label{fig:MagLL}
\end{figure}

A high-field Nb$_3$Sn accelerator magnet beyond the 16 T limit requires the use of HTS superconductor coils in the innermost layers, where the maximum field is achieved \cite{20TFerracin, 20TZlobin, 20TMarika}. For this reason, the first HTS stress-managed cosine-theta (SMCT) coil made of Bi-2212 Rutherford cable in dipole configuration, called Bi-SMCT1, was conceptualized at Fermilab and is currently under assembly at LBNL \cite{Bi2212InsertConc, Bi2212Model, Bi2212InsertDev1, Bi2212InsertDev2, Bi2212Tech}. This HTS insert coil is designed to fit the 60 mm aperture of the 2-layer standard cosine-theta Nb$_3$Sn demonstrator developed at Fermilab in the past years \cite{MDPCT1}. An 80 mm thick iron yoke surrounds the proposed design hybrid magnet in a dipole configuration. Its main parameters are reported in Table \ref{tab:ngeomelepar}.
The critical current parametrization at 4.2 K of the two superconductors is reported in Figure \ref{fig:MagLL}. The plot reports also the load lines of HTS insert and LTS outsert standalone powering, as well as the series connection powering. For each of those three powering configurations, the magnetic field in the bore (B$_O$) and the maximum magnetic field in the conductor (B$_C$) are plotted as a function of the energizing current. The series powering load-line intersects the Bi-2212 J$_C$ function at a current value of 9.63 kA and a bore field of 12.9 T.
Multiple magnetic simulations, with iron yoke, were performed in ANSYS APDL, varying I$_{HTS}$ and I$_{LTS}$ currents to obtain the magnetic field in the bore in different powering configurations. Those points were then interpolated to obtain the magnetic results in all the possible configurations.
The optimal currents for powering this specific hybrid magnet design were found to be I$_{HTS}$ = 9 kA and I$_{LTS}$ = 13.5 kA with approximately 16 T of bore field (see section \ref{sec:results}, Figure \ref{fig:MechResDeg}). The maximum values of the magnetic field in the bore and each conductor layer for both the series and optimal powering configurations are reported in Table \ref{tab:magres}.

\section{Mechanical Analysis}
\subsection{Mechanical Structure}
\begin{figure}
\centering
\includegraphics[width=3in]{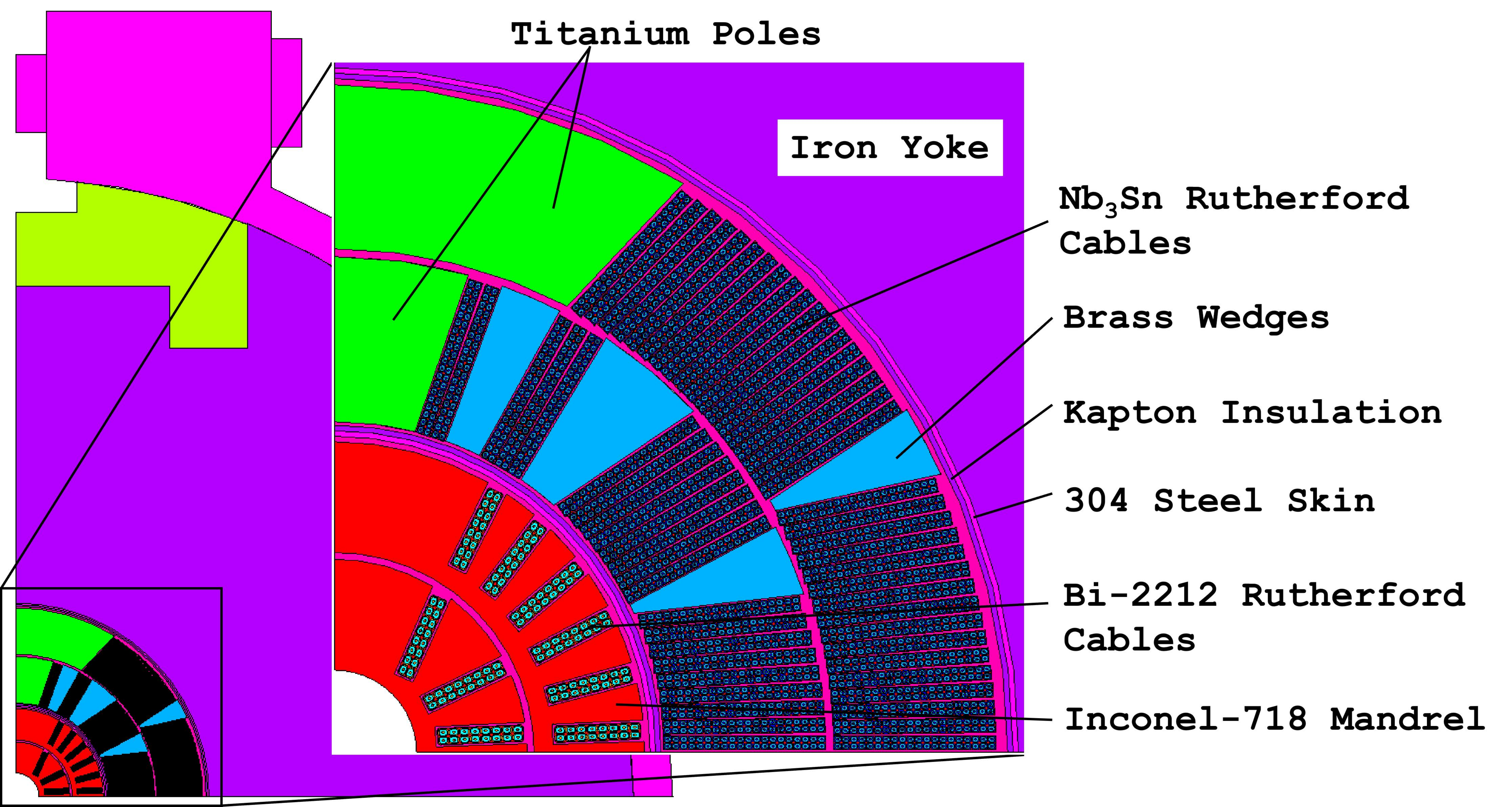}
\caption{Mechanical structure and materials of the hybrid dipole magnet implemented in ANSYS to compute the mechanical analysis.}
\label{fig:mechstruct}
\end{figure}

The mechanical simulation of the hybrid dipole modeled with the heterogeneous Rutherford cable model is computationally demanding. To simplify the analysis, the results of the entire mechanical structure, loaded at pre-stress, cool-down, and powering, were replaced with those of a simplified model shown in Figure \ref{fig:mechstruct}. The latter consists of removing the iron yoke, aluminum clamp, and stainless-steel shell from the simulation, inserting infinitely rigid boundary conditions at the dipole OD (yellow dotted line in Figure \ref{fig:mechstruct}), and loading the conductors with Lorentz forces only, considering exclusively the powering load step.


\subsection{Mechanical Results}\label{sec:results}

\begin{figure}
\centering
\includegraphics[width=3.4in]{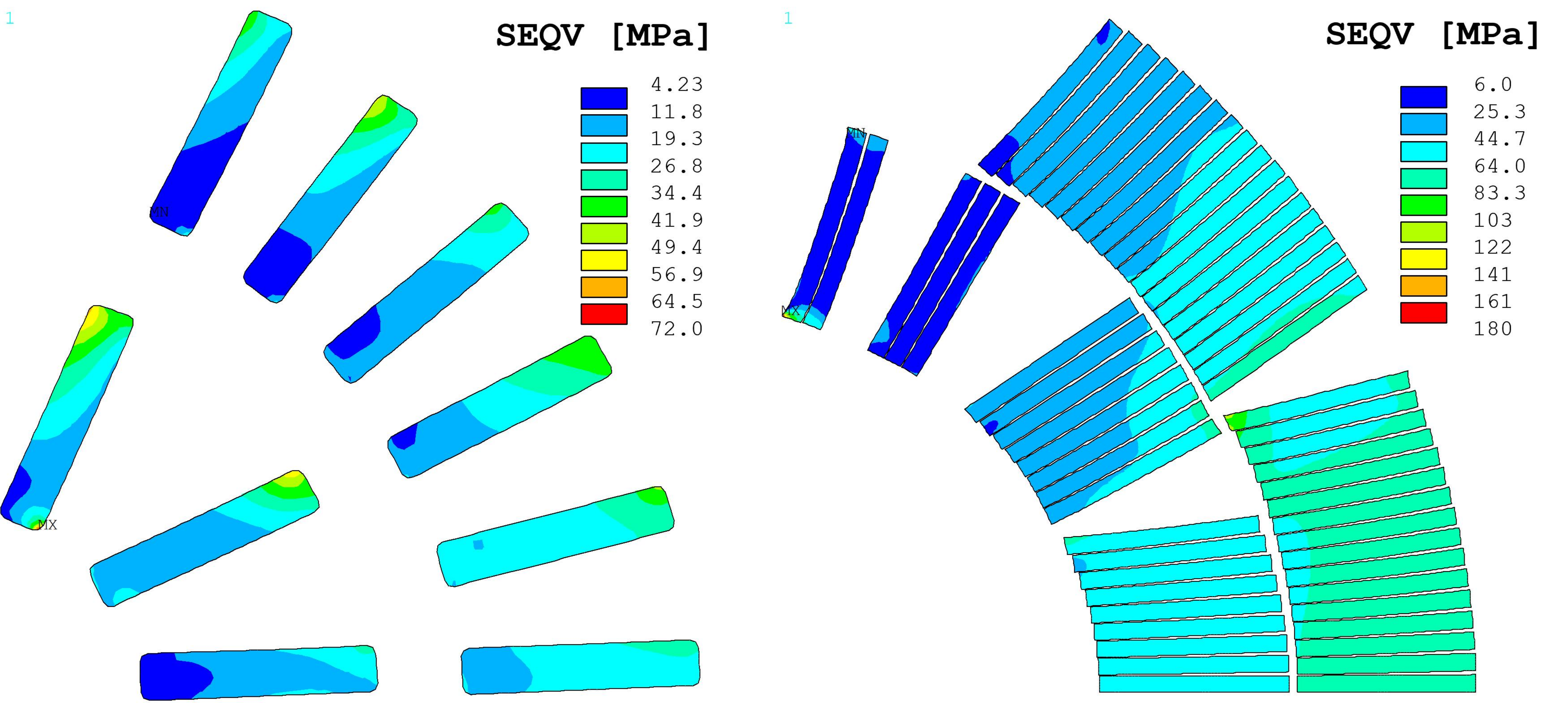}
\caption{Plot of the equivalent Von-Mises stress distribution in the HTS (left) and LTS (right) coil at optimal powering only (I$_{HTS}$ = 9 kA, I$_{LTS}$ = 13.5 kA), homogeneous cable model at 4.2 K.}
\label{fig:MechResSimpl}
\end{figure}

Figure \ref{fig:MechResSimpl} shows the equivalent Von Mises stress in the Rutherford cable areas computed in the mechanical analysis with the homogeneous cable model. The peak stress at powering in the Bi-2212 cable is 72 MPa, while in the Nb$_3$Sn is 215 MPa and localized at the corner of the second layer pole turn. The rest of the Nb$_3$Sn conductor is within the stress limit of 180 MPa.

\begin{figure}[bt]
\centering
\includegraphics[width=2.7in]{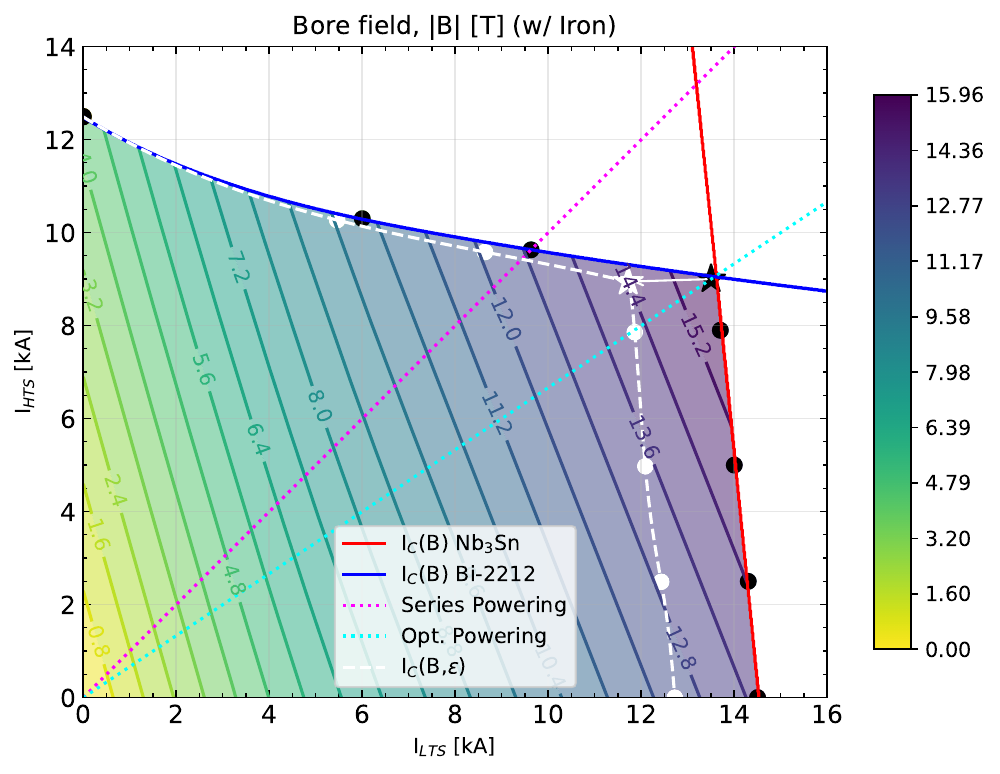}
\caption{Magnetic field in the bore of the hybrid dipole varying I$_{LTS}$ and I$_{HTS}$. The red and blue lines are the I$_C$(B) at zero strain for Nb$_3$Sn and Bi-2212, respectively. The black star represents the optimal powering, the white star the I$_C$($\varepsilon$) in that configuration. The white dotted line highlights the critical current degradation for all the configurations at 4.2 K.}
\label{fig:MechResDeg}
\end{figure}

The mechanical analysis of the hybrid magnet with heterogeneous Rutherford cable model enables the detailed computation of the superconducting areas' strain state distribution. The obtained $\varepsilon$ and $B$ arrays were used as input for the calculation of the critical current degradation I$_C$(B, T, $\varepsilon$) in each strand element of the two superconductors as defined in section \ref{sec:CCD}.
The final step of the analysis involved computing the critical current degradation of both superconductors for the entire set of powering configurations of the hybrid magnet at short sample limit. To do so, the magnetic field in the bore was plotted for all the combinations of I$_{HTS}$ and I$_{LTS}$ currents, highlighting the critical current limit for both the Bi-2212 and the Nb$_3$Sn at zero strain. Eight points on the critical surface were chosen, and for each of them, the mechanical simulations were computed on the LBNL Lawrencium cluster. For each run performed, results were collected as for the optimal powering configuration shown above. Results of this work are reported in Figure \ref{fig:MechResDeg}. At each black dot representing a specific powering configuration at zero strain, a white dot is associated, identifying the magnetic field with degraded critical current. The white dotted line shown in the plot is obtained by interpolating the computed results, and represents the conductors' I$_C$(B, T, $\varepsilon$) degradation for the powering configuration on their critical surface.

\begin{figure}[bt]
\centering
\includegraphics[width=3.4in]{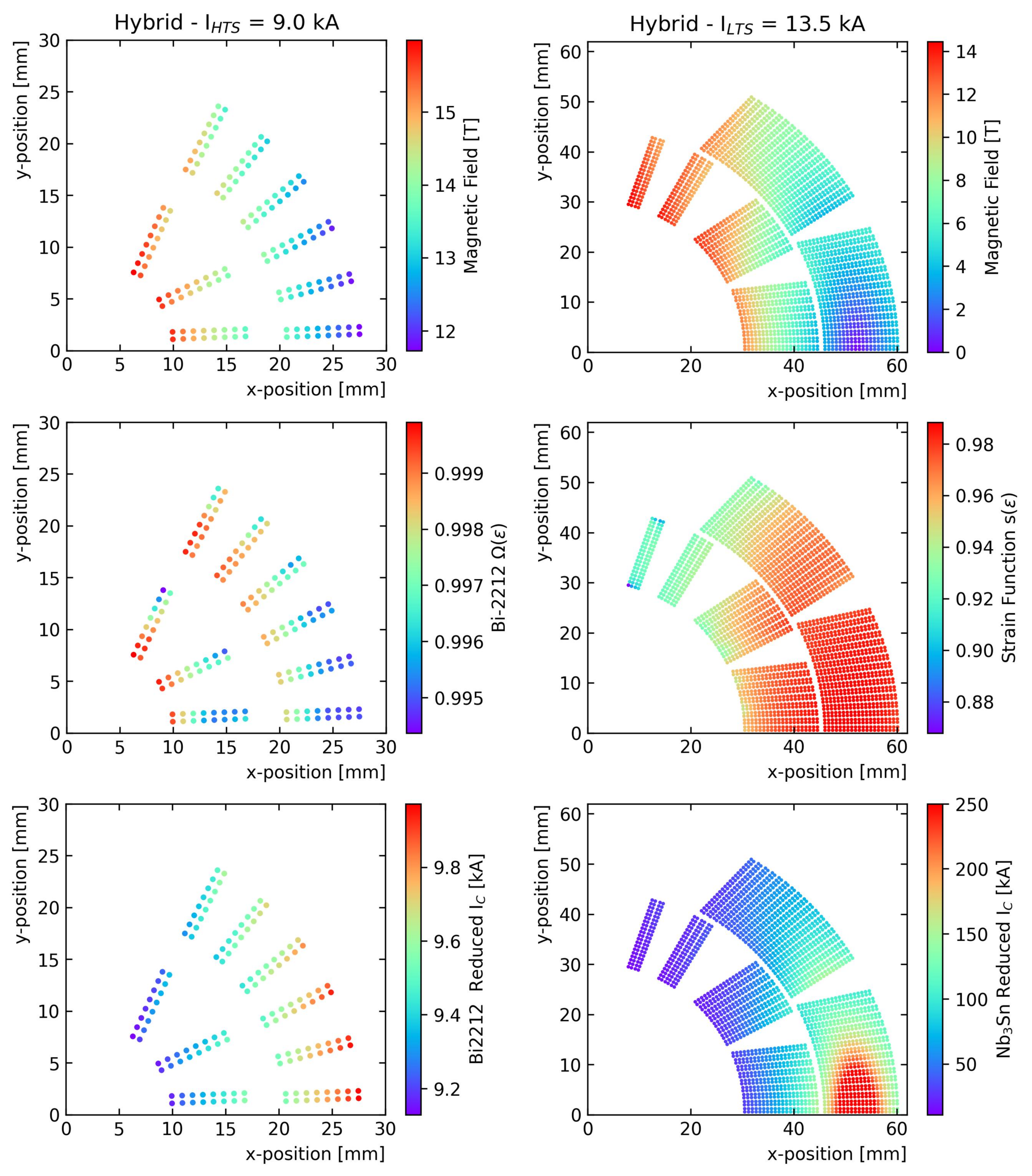}
\caption{Magnetic and mechanical analysis results for the optimal configuration: magnetic field in the conductor strands (top),  $\Omega$($\varepsilon$) and 'strain' function s($\varepsilon$) (center), and I$_C$(B, $\varepsilon$) current (bottom) in Bi-2212 strands (left) and Nb$_3$Sn strands (right).}
\label{fig:CCDEG}
\end{figure}

The peak of I$_C$ degradation in Bi-2212 and Nb$_3$Sn happens at optimal powering configuration, respectively with values of 0.5\% and 10\% (13\% due to stress intensification in the turn at pole region), and nominal I$_C$(B, T, $\varepsilon$) values of 9 kA and 14.9 kA. Figure \ref{fig:CCDEG} reports the results obtained for the optimal powering configuration of the hybrid magnet, specifically the magnetic field in the conductors, the $\Omega$($\varepsilon$) for the Bi-2212, the strain function for the Nb$_3$Sn, and the degraded I$_C$(B, T, $\varepsilon$) in each strand of the conductors.
In the degraded configuration, the hybrid dipole can generate a bore field of 14.4 T, which is 9.8\% lower than the bore field at zero strain. The relatively small I$_C$ reduction in the Bi-2212 Rutherford cable compared to the one in the LTS is due to the effect of the SMCT design for the HTS insert, where stresses are progressively distributed on the mandrel structure.

\section{Conclusion}
In this paper, the magnetic analysis of the hybrid HTS/LTS dipole developed at FNAL is presented for every combination of powering currents. The magnetic results of the series and optimal configurations at the short sample limit were reported and compared. For these two configurations, a mechanical analysis using the heterogeneous Rutherford cable model was performed to evaluate the critical current reduction due to strain during powering.
The distribution of the magnetic field in the conductor, the critical current degradation, and the nominal degraded current in each strand were reported for the optimal configuration.

The computation of the I$_C$(B, T, $\varepsilon$) for both Bi-2212 and Nb$_3$Sn was performed for all the powering configurations at short sample limit to investigate the behavior of the HTS/LTS superconductor within the hybrid dipole structure. Results show that with the chosen parametrization, the Bi-2212 critical current reduction due to strain is lower than the Nb$_3$Sn I$_C$ reduction. This underlines the conductor benefits of the SMCT structure implemented for the Bi-2212 insert compared to the standard cosine-theta design.
The reported procedure utilizing the heterogeneous Rutherford cable model can be extended to alternative hybrid magnet designs and conductor shapes. 


\section*{Acknowledgments}
The author wants to thank Lucas Brower for the valuable technical advice given to enhance the quality of this work.

\vfill


\begin{thebibliography}{1}
\bibliographystyle{IEEEtran}

\bibitem{USMDP}
The U.S. Magnet Development Program. "\url{https://usmdp.lbl.gov/strategic-documents}".

\bibitem{LTS1}
A.V. Zlobin, J.R. Carmichael, V.V. Kashikhin, I. Novitski, "Conceptual Design of a 17 T Nb$_3$Sn Accelerator Dipole Magnet", 9th International Particle Accelerator Conference, IPAC'18, doi:10.18429/JACoW-IPAC2018-WEPML027.

\bibitem{LTS2}
I. Novitski, A. V. Zlobin, J. Coghill, E. Barzi and D. Turrioni, "Development of a 120-mm Aperture Nb$_3$Sn Dipole Coil With Stress Management," in IEEE Transactions on Applied Superconductivity, vol. 32, no. 6, pp. 1-5, Sept. 2022, Art no. 4006005, doi: 10.1109/TASC.2022.3163062. \url{https://ieeexplore.ieee.org/document/9756214}

\bibitem{LTS3}
I. Novitski, A. V. Zlobin, E. Barzi and D. Turrioni, "Design and Assembly of a Large-Aperture Nb$_3$Sn Cos-Theta Dipole Coil With Stress Management in Dipole Mirror Configuration," in IEEE Transactions on Applied Superconductivity, vol. 33, no. 5, pp. 1-5, Aug. 2023, Art no. 4001405, doi: 10.1109/TASC.2023.3244894. \url{https://ieeexplore.ieee.org/document/10057115}

\bibitem{LTS4}
A. V. Zlobin et al., "Development and test of a large-aperture Nb$_3$Sn cos-theta dipole coil with stress management", IPAC-24. \url{https://doi.org/10.18429/JACoW-IPAC2024-WEPS68}

\bibitem{LTS5}
A. V. Zlobin, I. Novitski, M. Baldini and E. Barzi, "Quench Performance of a Large-Aperture Nb$_3$Sn Cos-Theta Coil With Stress Management in Dipole Mirror Configurations," in IEEE Transactions on Applied Superconductivity, vol. 35, no. 6, pp. 1-5, Sept. 2025, Art no. 0600505, doi: 10.1109/TASC.2024.3508572. \url{https://ieeexplore.ieee.org/document/10816553}









\bibitem{GVmethod}
G. Vallone, E. Anderssen, B. Bordini, P. Ferracin, J. F. Troitino and S. Prestemon, "A methodology to compute the critical current limit in Nb$_3$Sn magnets", 2021 Supercond. Sci. Technol. 34 025002 \url{https://iopscience.iop.org/article/10.1088/1361-6668/abc56b}

\bibitem{ADmodel}
A. D'Agliano et al., "FEM Analysis of Hybrid LTS/HTS Cos-Theta Dipole Magnet With Heterogeneous Cable Model," in IEEE Transactions on Applied Superconductivity, vol. 35, no. 5, pp. 1-5, Aug. 2025, Art no. 4000105, doi: 10.1109/TASC.2024.3502578. \url{https://ieeexplore.ieee.org/document/10758385}

\bibitem{CCDGiorgio}
G. Vallone, B. Bordini and P. Ferracin, "Computation of the Reversible Critical Current Degradation in Nb$_3$Sn Rutherford Cables for Particle Accelerator Magnets," in IEEE Transactions on Applied Superconductivity, vol. 28, no. 4, pp. 1-6, June 2018, Art no. 4801506, doi: 10.1109/TASC.2018.2810222. \url{https://ieeexplore.ieee.org/document/8303782}

\bibitem{FNALmodel}
E. Barzi, C. Franceschelli, I. Novitski, F. Sartori and A. V. Zlobin, "Measurements and Modeling of Mechanical Properties of Nb$_3$Sn Strands, Cables, and Coils," in IEEE Transactions on Applied Superconductivity, vol. 29, no. 5, pp. 1-8, Aug. 2019, Art no. 8401808, doi: 10.1109/TASC.2019.2906734. \url{https://ieeexplore.ieee.org/document/8681728}

\bibitem{Irreversible}
Patrick Ebermann et al., "Irreversible degradation of Nb$_3$Sn Rutherford cables due to transverse compressive stress at room temperature", 2018 Supercond. Sci. Technol. 31 065009. \url{https://iopscience.iop.org/article/10.1088/1361-6668/aab5fa}




\bibitem{CCDBordini}
B. Bordini, P. Alknes, L. Bottura, L. Rossi and D. Valentinis, "An exponential scaling law for the strain dependence of the Nb$_3$Sn critical current density", 2013, Supercond. Sci. Technol., 26, 075014, DOI 10.1088/0953-2048/26/7/075014. \url{https://iopscience.iop.org/article/10.1088/0953-2048/26/7/075014}

\bibitem{CCDBi2212}
A. Kario et al., "Critical Current Reduction by Transverse Pressure in Bi-2212 Rutherford cables up to 300 MPa at 11 T / 4.2 K", HFM Annual Meeting, Feb. 2025, CERN. \href{chrome-extension://efaidnbmnnnibpcajpcglclefindmkaj/https://indico.cern.ch/event/1471305/contributions/6260702/attachments/3013367/5313673/Critical%20Current%20Reduction%20by%20Transverse%20Pressure%20in%20Bi-2212.pdf}{online}.



\bibitem{Areview}
G. Vallone, E. Anderssen, B. Bordini and P. Ferracin, "A Review of the Mechanical Properties of Materials Used in Nb$_3$Sn Magnets for Particle Accelerators," in IEEE Transactions on Applied Superconductivity, vol. 33, no. 5, pp. 1-6, Aug. 2023, Art no. 4002806, doi: 10.1109/TASC.2023.3248544. \url{https://ieeexplore.ieee.org/document/10050867}





\bibitem{20TFerracin}
P. Ferracin et al., "Towards 20 T Hybrid Accelerator Dipole Magnets," in IEEE Transactions on Applied Superconductivity, vol. 32, no. 6, pp. 1-6, Sept. 2022, Art no. 4000906, doi: 10.1109/TASC.2022.3152715. \url{https://ieeexplore.ieee.org/document/9718198}

\bibitem{20TZlobin}
A. V. Zlobin and I. Noviski and E. Barzi and P. Ferracin, "20 T Dipole Magnet Based on Hybrid HTS/LTS Cos-Theta Coils with Stress Management," 2023, , 2305.06776. \url{https://arxiv.org/abs/2305.06776}

\bibitem{20TMarika}
M. D’Addazio, P. Ferracin, Senior Member, IEEE, V. Marinozzi, E. Ravaioli, G. Vallone, and L. Savoldi, Member, IEEE, "Conceptual Structural Design and Analysis of a 20 T Hybrid Cos-Theta Dipole for Future Particle Colliders", IEEE Transactions on Applied Superconductivity, Vol. 35, No. 5, August 2025. \url{https://ieeexplore.ieee.org/document/10786366}





\bibitem{Bi2212InsertConc}
A. V. Zlobin, I. Novitski, E. Barzi, "Conceptual Design of a HTS Dipole Insert Based on Bi2212 Rutherford Cable," in Instruments 4, no. 4, 29. \url{https://doi.org/10.3390/instruments4040029}

\bibitem{Bi2212Model}
A. D’Agliano et al., "Magnetic and mechanical analysis of Bi-2212 Rutherford cable in a cos-theta sub-scale dipole coil", 2025 Supercond. Sci. Technol., 38, 035021. \url{https://iopscience.iop.org/article/10.1088/1361-6668/adb340}

\bibitem{Bi2212InsertDev1}
A. V. Zlobin, I. Novitski, E. Barzi, and D. Turrioni, "Development of a Small-Aperture Cos-Theta Dipole Insert Coil Based on Bi2212 Rutherford Cable and Stress Management Structure", IEEE Transactions on Applied Superconductivity, vol. 32, no. 6, pp. 1-5, Sept. 2022, Art no. 4003605. \url{https://ieeexplore.ieee.org/document/9734728}

\bibitem{Bi2212InsertDev2}
A. V. Zlobin, I. Novitski, E. Barzi, and D. Turrioni, "Development of a Bi2212 Dipole Insert at Fermilab," in IEEE Transactions on Applied Superconductivity, vol. 33, no. 5, pp. 1-5, Aug. 2023, Art no. 4602305. \url{https://ieeexplore.ieee.org/document/10092279}

\bibitem{Bi2212Tech}
E. Barzi, "Conductor Properties and Coil Technology for a Bi2212 Dipole Insert for 20 Tesla Hybrid Accelerator Magnets", Contribution to Snowmass 2021, Oct. 2022, \url{https: //doi.org/10.48550/arXiv.2204.01072}


\bibitem{MDPCT1}
A. V. Zlobin et al., "Development and First Test of the 15 T Nb$_3$Sn Dipole Demonstrator MDPCT1," in IEEE Transactions on Applied Superconductivity, vol. 30, no. 4, pp. 1-5, June 2020, Art no. 4000805, doi: 10.1109/TASC.2020.296768. \url{https://ieeexplore.ieee.org/document/8989965}





\end{thebibliography}
\end{document}